\documentclass[preprint2]{aastex}

\usepackage{color}

\shorttitle{Abundance of massive subhalos}
\shortauthors{Munari E., et al.}

\begin{document}


\title{Numerical simulations challenged on the prediction of massive subhalo abundance in galaxy clusters: the case of Abell 2142}


\author{E. Munari\altaffilmark{1,2,3}, C. Grillo\altaffilmark{1,4}, G. De Lucia\altaffilmark{2}, A. Biviano\altaffilmark{2} , M. Annunziatella\altaffilmark{2}, S. Borgani\altaffilmark{2,3}, M. Lombardi\altaffilmark{4}, A. Mercurio\altaffilmark{5}, P. Rosati\altaffilmark{6} }
\email{munari@oats.inaf.it}

\altaffiltext{1}{Dark Cosmology Centre, Niels Bohr Institute, University of Copenhagen, Juliane Maries Vej 30, DK-2100 Copenhagen, Denmark}
\altaffiltext{2}{INAF - Osservatorio Astronomico di Trieste, via G. B. Tiepolo 11, I-34143, Trieste, Italy}
\altaffiltext{3}{Dipartimento di Fisica, Universit\`a degli Studi di Trieste, via G. B. Tiepolo 11, I-34143 Trieste, Italy}
\altaffiltext{4}{Dipartimento di Fisica, Universit\`a degli Studi di Milano, via Celoria 16, I-20133 Milano, Italy}
\altaffiltext{5}{INAF - Osservatorio Astronomico di Capodimonte, Via Moiariello 16, I-80131 Napoli, Italy}
\altaffiltext{6}{Dipartimento di Fisica e Scienze della Terra, Universit\`a degli Studi di Ferrara, Via Saragat 1, I-44122 Ferrara, Italy}


\begin{abstract}
In this Letter we compare the abundance of member galaxies of a rich,
nearby ($z=0.09$) galaxy cluster, Abell 2142, with that of halos of
comparable virial mass extracted from sets of state-of-the-art
numerical simulations, both collisionless at different resolutions and
with the inclusion of baryonic physics in the form of cooling, star
formation, and feedback by active galactic nuclei. We also use two
semi-analytical models to account for the presence of orphan
galaxies. The photometric and spectroscopic information, taken from
the Sloan Digital Sky Survey Data Release 12 (SDSS DR12) database,
allows us to estimate the stellar velocity dispersion of member
galaxies of Abell 2142. This quantity is used as proxy for the total
mass of secure cluster members and is properly compared with that of
subhalos in simulations. We find that simulated halos have a
statistically significant ($\gtrsim 7$ sigma confidence level) smaller
amount of massive (circular velocity above $200\,{\rm km\, s^{-1}}$)
subhalos, even before accounting for the possible incompleteness of
observations. These results corroborate the findings from a recent
strong lensing study of the Hubble Frontier Fields galaxy cluster MACS
J0416 \citep{grillo2015} and suggest that the observed difference is
already present at the level of dark matter (DM) subhalos and is not
solved by introducing baryonic physics. A deeper understanding of this
discrepancy between observations and simulations will provide valuable
insights into the impact of the physical properties of DM particles
and the effect of baryons on the formation and evolution of
cosmological structures.
\end{abstract}





\section{Introduction}
\label{sect: intro}
In the hierarchical structure formation scenario, galaxies form and
evolve in dark matter (DM hereafter) halos that merge with other halos
to assemble larger systems. Because of this process, halos are
composed of a diffuse matter component and a population of subhalos,
whose motion and spatial distribution is determined by dynamical
processes taking place after a halo has merged into another
one. Dynamical friction makes subhalos sink toward the halo center,
where strong tidal fields are very effective at stripping material
from the external regions of subhalos \citep[see,
  e.g.][]{kravtsov2004,BK2008}. Several other processes can act and
affect the DM and baryonic components, such as tidal heating,
ram-pressure stripping and harassment
\citep[e.g.][]{moran2007,biviano08,bruggen2008,delucia2012}.

In \cite{grillo2015}, the subhalo distribution inferred from the
strong lensing analysis of a massive galaxy cluster, MACS J0416, is
compared with the predictions of N-body simulations. This comparison
shows a significant lack of massive subhalos in simulations. The
latter do not include baryonic physics and this could be one of the
reasons for the disagreement with the observed subhalo population. In
fact, the simulated subhalos are less concentrated than they would be
if they had baryons, therefore they are more fragile against tidal
stripping. On the other hand, that cluster has a total density profile
characterized by an inner core. This would cause tidal fields to be
weaker than in the case of a cuspy profile, such as the
Navarro-Frenk-White profile \citep{navarro96,navarro97} found in
simulated halos, leading to a larger massive subhalo population in
MACS J0416 when compared with simulations.

In this paper, we analyze the subhalo distribution of a massive,
nearby cluster, by utilizing internal kinematics of cluster galaxies
as a proxy of subhalo masses, as opposed to the strong lensing
modeling techniques used in \cite{grillo2015}. We then compare the
observed subhalo population with the predictions of numerical
simulations.

In particular, we study Abell 2142 (A2142 hereafter), a massive ($\rm
M_{200,cr} = (1.25 \pm 0.13) \times 10^{15}\rm M_\odot$) cluster at $z
\sim 0.09$ \citep[][M14 hereafter]{munari14}. The cluster was studied
by several authors using different probes, namely X-ray
\citep{markevitch00,akamatsu11,rossetti13}, the Sunyaev-Zel'dovich effect
\citep{umetsu09}, weak lensing \citep{okabe08} and galaxy dynamics
(\citealt[][O11 hereafter]{owers11}; M14), and although it possesses
several subclumps in the galaxy distribution (O11), these do not
appear to affect the dynamical equilibrium of the cluster
significantly (M14).

\section{The data set}

\subsection{The observations}
\label{sect: observations}
Based on the spectroscopic catalog provided by O11, here we use the
membership of A2142 that was computed by M14. We restrict our analysis
to the inner 2.2 Mpc in projection, which is very close to the virial
radius of A2142 found by M14, and, as done in that study, adopt the X-ray
center provided by \cite{degrandi02} as the cluster center. The number
of galaxy members extracted from the O11 catalog is $N_{\rm mem} =
721$ within 2.2 Mpc from the cluster center. The analysis by O11 has a
magnitude limit of 20.5 in the $R$ band of the Johnson-Cousins system
($R_{\rm JC}$ hereafter). We anticipate that this limit corresponds to
values of circular velocity that are well below those that we probe in
this work, thus not affecting our main results.

We use the photometric information contained in the Sloan Digital Sky
Survey Data Release 12 (SDSS DR12)
database\footnote{\url{http://www.sdss.org/dr12/}}, selecting galaxies
with $238\fdg983 < \rm{R.A.} < 240\fdg183$, $ 26\fdg633 < \rm{Dec.} <
27\fdg834$, and $\rm{petroMag_r} < 25$. This magnitude limit ensures
that we include all the members. We match this catalog with the member
catalog by requiring objects to be closer than 0.06 arcsec ($\simeq
6\, \rm kpc$ at the cluster redshift) in projection. The number of
cluster members with SDSS photometric information is $N_{\rm mem \cap
  ph} = 708$, within 2.2 Mpc from the cluster center.

Then we extract from the SDSS DR12 spectroscopic sample the galaxies
that satisfy the same criteria used for the photometric selection,
further requiring only objects with secure redshifts, and retrieve
$N_{\rm sp} = 288$ galaxies within 2.2 Mpc from the cluster
center. For these objects, we estimate aperture-corrected stellar
velocity dispersion values $\sigma_0$, i.e. the velocity dispersion of
stars within an eighth of the galaxy effective radius, following the
prescription presented in \cite{jorgensen95}, and using the SDSS
values of the galaxy effective radii $R_{e,i}$ in the $i$ band. We
perform the matching of this catalog with that of cluster members by
imposing the same limit in projected distance as before and a relative
difference in redshift smaller than $1\%$. The matched cluster
galaxies are $N_{\rm mem \cap sp} = 187$.

We further restrict our analysis to elliptical galaxies by considering
only objects with $\verb|fracDev_i| > 0.8$. This quantity is the best
fitting coefficient of the de Vaucouleurs term obtained from the
decomposition of the surface brightness profile in the $i$ band of
each galaxy in terms of a linear combination of exponential and de
Vaucouleurs profiles. This photometric criterion ensures that the
selected objects are very likely elliptical galaxies \citep[see
  also][]{grillo2010}. The number of elliptical members is $N_{\rm mem
  \cap sp,E} = 146$.

According to the same criterion, we also select elliptical galaxies
from the photometric catalog, with the further requirement that
$R_{e,i} > 0.3\,\rm kpc$, to exclude galaxies with unreliable
effective radii. This results in $N_{\rm mem \cap ph,E} = 324$
galaxies within 2.2 Mpc from the cluster center.

\subsection{The simulations}
\label{sect: simulations}
Here we briefly summarize the main features of the simulations used
for this work and we refer the reader to \cite{rasia15} for a more
detailed description. Starting from a low resolution DM-only
simulation, 29 massive clusters are identified and resimulated at
higher resolution with the zoom-in technique, with different
implementations for baryonic physics. The low resolution parent
simulation consists of $1024^3$ particles in a $1\,{\rm h^{-1}Gpc}$
box, realized with the {\sc GADGET 3} code, an improved version of the
{\sc GADGET 2} code \citep{springel2005}. A flat $\Lambda \rm CDM$
cosmology with $\Omega_m=0.24$, $\Omega_{bar}=0.04$, $H_0=72\, {\rm
  km\,s^{-1}\,Mpc^{-1}}$, $n_s=0.96$ and $\sigma_8=0.8$ is adopted. In
this work we use four sets of such simulations, with two different
implementations of baryonic physics. The ``DMHR'' is a collisionless
realization with a particle mass of $m_{\rm DM} = 1 \times 10^8\, {\rm
  h^{-1}M_\odot}$ and a Plummer-equivalent softening length of
$\epsilon = 2.5\,\rm{h}^{-1}$ kpc in physical units below z = 2 (fixed
in comoving units at higher redshift). The ``DMLR'' is a
lower-resolution version of the DMHR, with $m_{\rm DM} = 1 \times
10^9\, {\rm h^{-1}M_\odot}$ and $\epsilon = 5\,\rm{h}^{-1}$ kpc. The
``CSF'' set implements a metallicity-dependent radiative cooling and a
sub-resolution model for star formation with Chabrier IMF
\citep{chabrier03}. A uniform time-dependent UV background is
included, while kinetic feedback contributed by supernovae is
implemented in the form of winds with a velocity of $\sim 350\, {\rm
  km\, s^{-1}}$. Metal production and ejection into the inter-stellar
medium are modeled as in \cite{tornatore07}. The ``AGN'' set
implements the additional effect of active galactic nuclei feedback,
modeled following \cite{steinborn15}. The two sets with baryon physics
have the same DM particle mass and force resolution as the DMLR
set. The algorithm SUBFIND \citep{springel2001,dolag2009} is used to
separate the subhalos from the diffuse matter of the cluster halo, by
searching overdensities of bound DM and star (in the hydrodynamical
runs) particles in the cluster density field.

The baryonic component of a subhalo, i.e. a galaxy, is usually more
compact than the subhalo, and is therefore more resistant against the
stripping due to cluster tidal fields. The external part of a subhalo
can eventually be entirely stripped, leaving only the galaxy. Such
galaxies, that have lost their DM envelope, are called
``orphans''. N-body simulations, lacking the compact baryonic cores,
are not able to reproduce the population of orphan galaxies. On the
contrary, hydrodynamical simulations have the compact baryonic cores,
but due to resolution limitations, they might not capture the whole
population of orphan galaxies. For this reason, we also use two
different semi-analytical models, namely those described in
\citet[][hereafter DLB07]{DLB07} and \citet[][hereafter HEN15]{HEN15},
to which we refer for further details. Both these models use subhalo
merger trees extracted from high-resolution N-body simulations (the
``Millennium'' simulation \citep{springel2005}, on which subhalos are
identified using SUBFIND as in the hydrodynamical simulations
described above) as input. When a subhalo is stripped below the
resolution of the simulation, the galaxy it hosts becomes an `orphan'
galaxy and is assigned a residual merger time based on variations of
the classical Chandrasekhar formula \citep{chandrasekhar1943}. The two
models differ for the specific prescriptions adopted to describe
various physical processes, and predict different amounts of orphan
galaxies in clusters.

We select galaxy clusters with virial masses values larger than
$10^{15}\rm M_\odot$ at the redshift closest to that of A2142, namely
$z=0$ for DMHR and DMLR, $z=0.1$ for CSF and AGN, $z=0.09$ for DLB07,
and $z=0.08$ for HEN15. In this way we select galaxy clusters with
virial masses comparable to that of A2142, namely $1.25 \times
10^{15}\rm M_\odot$ (M14). We simulate the line-of-sight effects of
observations by projecting each cluster along three orthogonal
directions. The selected samples consist of 22, 25, 22, 21, 18, 23
clusters, observed in 3 directions, for a total of 66, 75, 66, 63, 54,
69 systems in projection in the DMHR, DMLR, CSF, AGN, DLB07, HEN15
sets, respectively. We consider the two-dimensional (2D) distance of
each subhalo from the cluster center, selecting those within 2 virial
radii along the direction of projection, in order to include objects
residing in the cluster outskirts, but excluding interlopers. For all
subhalos, we store the values of circular velocity $v_c$, defined as
the maximum value of $\sqrt{G\, M(<r)/r}$, $M(<r)$ being the mass
within the three-dimensional (3D) distance $r$ from the center of the
subhalo. For the orphan galaxies in the semi-analytic models, the
adopted circular velocity is the value the subhalo had at the last
snapshot before disappearing below the resolution limit.

\section{Circular velocity distribution}
\label{sect: vc}

Several observational and theoretical studies have shown that in
massive early-type galaxies (ETGs) the DM and stellar components
combine to produce a total mass density profile that can be
well-approximated by an isothermal profile, although neither of the
two components have precisely such a profile. This is the so-called
``bulge-halo conspiracy''. ETGs from the SLACS survey were studied by
\cite{auger2010}, combining different probes, and by
\cite{barnabe2011} combining stellar kinematics and gravitational
lensing, while \cite{cappellari2015} applied dynamical models on a
sample of 14 fast rotators ETGs out to a median radius of $\sim 4 \,
R_e$ ($\sim \rm 10\,kpc$). All these studies found that an isothermal
profile is a good description of the galaxy total mass density
profile. Using the \emph{Chandra X-ray observatory},
\cite{humphrey2010} studied a sample of objects spanning $\sim2$
orders of magnitude in mass, from ETGs to galaxy clusters. They
concluded that an isothermal profile is a good description of the
total mass profile of galaxies out to $\sim10\,R_e$, where DM
dominates the mass budget. \cite{gavazzi2007} performed a joint strong
and weak lensing analysis of 22 massive ETGs, finding that the lenses
are described well by an isothermal profile out to $\sim 100\, R_e$
(few hundreds kpc), with an effective velocity dispersion value very
similar to that of the central stellar velocity dispersion.

On the theoretical side, \cite{dutton2014} constructed $\Lambda \rm
CDM$-based mass models to reproduce the observed structural and
dynamical scaling relations of ETGs in the SDSS and showed that all
models produce roughly isothermal total mass density profiles. By
studying 35 simulated spheroidal galaxies, \cite{remus2013} concluded
that the isothermal profile, resulting from the combination of the
stellar and DM components, acts as an attractor solution of the
complex dynamics of the system (although this mechanism has still
unclear explanations).
 
In the innermost regions of ETGs, stars dominate the total mass budget
and so here their velocity dispersion $\sigma_0$ is representative of
the velocity dispersion of the whole system, as shown, e.g. in
\cite{saglia1992} and \cite{thomas2007}. \cite{treu2006} and
\cite{grillo2008} have found that $\sigma_0$ is, within the
uncertainties, equal to $\sigma_{1D}$, which is the parameter that
characterizes an isothermal profile. On the other hand, in simulations
it is straightforward to measure the circular velocity of a subhalo,
$v_c$, as explained in Sect. \ref{sect: simulations}.

For a generic system, the 1-dimensional velocity dispersion $\sigma$
and circular velocity $v_c$ are related as follows:
$v_c=\sigma\times\sqrt{\gamma}$ where $\gamma=-d\ln\rho / d\ln r$ is
the logarithmic derivative of the mass density. If we assume that the
stellar mass density profile of ETGs is well described by their
luminosity distribution, and use the Jaffe model \citep{jaffe1983} for
it, we obtain $\gamma=2$ at $r=0$, and $\gamma=4$ at large
r. Hereafter we use $\gamma=2$; using $\gamma>2$ would only strengthen
our conclusions (see below). A conversion factor close to $\sqrt{2}$
or slightly higher is reported in \cite{cappellari2013}, who performed
a detailed axisymmetric dynamical modeling of a large sample of
observed ETGs. In the following, the values of velocity dispersion
will be converted into circular velocity values to compare
observations against simulations.

\begin{figure}
  \includegraphics[width=\columnwidth]{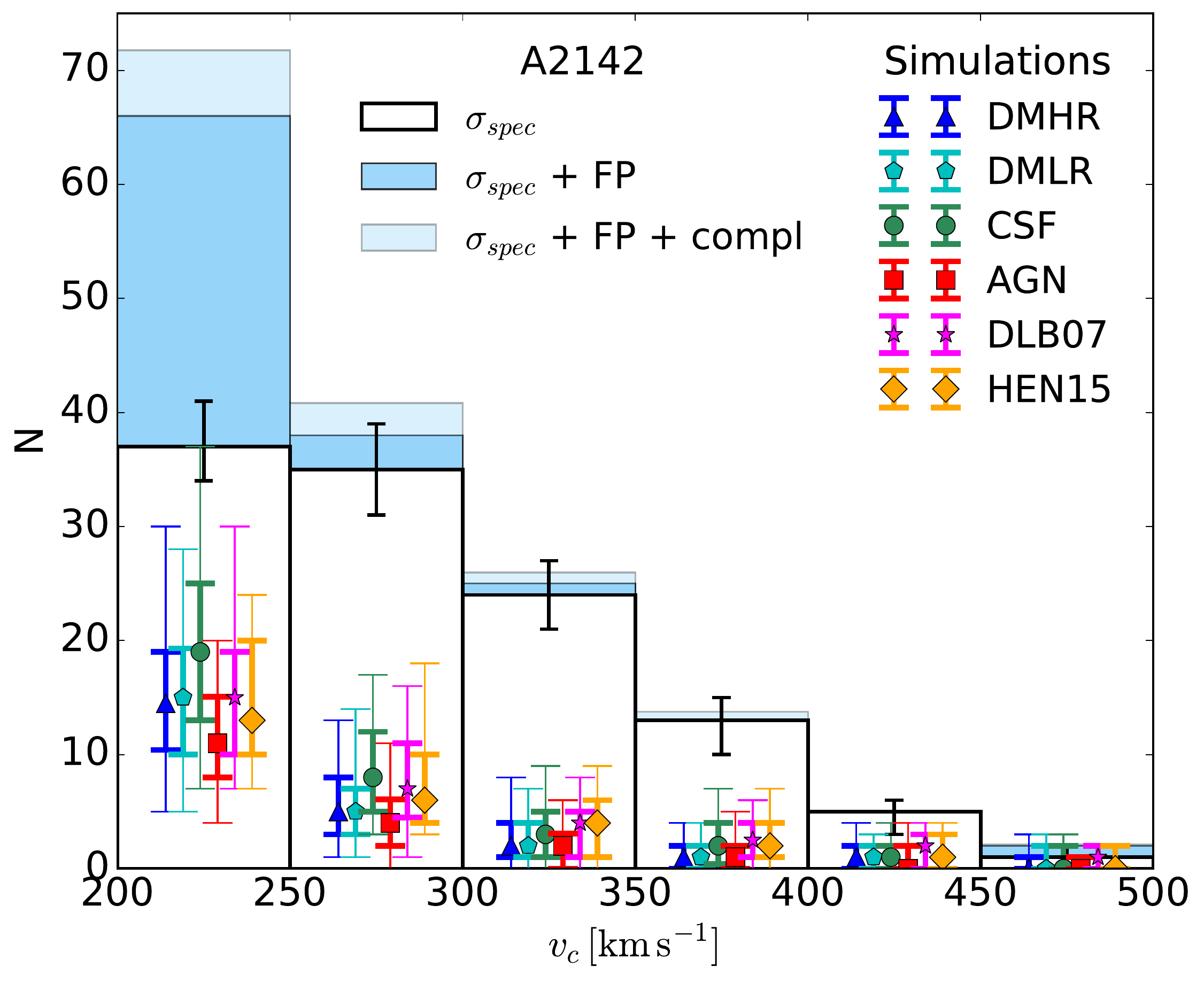}
  \caption{\label{fig: NofV} Distribution of the values of circular
    velocity of member galaxies within 2.2 Mpc in projection from the
    cluster center. The white histogram refers to the sample of A2142
    members with measured velocity dispersion from the SDSS DR12
    spectroscopic sample ($\rm mem \cap sp,E$). The histogram with
    error bars represents the median value, with its 16th and 84th
    percentiles, in each bin (see the text for details). The blue
    histogram extends the distribution to members that have velocity
    dispersions estimated from the Fundamental Plane. When the sample
    completeness is considered, the pale blue histogram is
    obtained. The symbols with error bars are the median values of
    circular velocity of subhalos in different simulated clusters, as
    indicated in the legend. Thick and thin error bars indicate the
    16th-84th percentiles and the minimum-maximum values,
    respectively.}
\end{figure}

In Figure \ref{fig: NofV}, we show the distribution of values of
circular velocity of A2142 member elliptical galaxies, within 2.2 Mpc
from the cluster center. The white histogram refers to the sample of
elliptical cluster members with measured velocity dispersion from the
SDSS spectroscopic sample ($\rm mem \cap sp,E$). To account for the
errors on the measurement of the velocity dispersion, we consider
10,000 realizations of the sample, where the value of the velocity
dispersion of each galaxy is taken from a normal distribution having
mean and standard deviation values that are equal, respectively, to
those of the aperture-corrected SDSS central stellar velocity
dispersion and its error. The histogram shows the median value in each
bin, and the error bars represent the 16th and 84th percentiles.

Then, we use the elliptical cluster members that have measured values
of velocity dispersion $\sigma_0$ ($\rm mem \cap sp,E$) to calibrate
the Fundamental Plane and estimate the velocity dispersion values of
the elliptical cluster members that do not have SDSS spectroscopic
information. The Fundamental Plane is a scaling law \citep[see,
  e.g.,][and references
  therein]{djorgovski1987,dressler1987,bernardi2003} that relates the
values of effective radius, central stellar velocity dispersion, and
surface brightness $\rm SB_e$ within the effective radius of
elliptical galaxies:
\begin{equation}
{\rm Log}(\sigma_0) = \alpha + \beta  \times {\rm Log}(R_e) - \gamma \times {\rm SB_e} 
\end{equation}
We restrict our analysis to the galaxies with velocity dispersion
above $90\, {\rm km\,s^{-1}}$ and find the following values: $\alpha =
6.84\pm0.26$, $\beta = 0.64\pm0.04$, $\gamma = 0.24\pm0.01$.  From
this relation, we get an estimate of the velocity dispersion of the
elliptical members for which a spectroscopic measurement is not
available. The blue histogram of Figure \ref{fig: NofV} shows the
distribution of the values of circular velocity of the members that
 have either a spectroscopic measurement of velocity dispersion or a
velocity dispersion estimate inferred from the Fundamental Plane, as
explained above. When this last histogram is corrected to account for
the incompleteness of the sample, we obtain the pale blue
histogram. We here provide an approximate estimate of the
completeness. O11 provide an estimate of the completeness of their
spectroscopic catalog by comparing it with their photometric catalog
(see their Figure 2). We use their estimate as completeness $C_M$. We
consider both the radial and the magnitude dependence of $C_M$. Since
O11 use the Johnson-Cousins R system, we convert the $g$ and $r$ SDSS
bands into that system to account for the magnitude dependence of
$C_M$. To each galaxyq in the sample of member elliptical
galaxies ($\rm mem \cap ph,E$), we assign a weight equal to
$C_M^{-1}$, according to the galaxy projected distance from the
cluster center and magnitude. In this way, we compute the
completeness-corrected distributions.

The symbols with error bars in Figure \ref{fig: NofV} are the median
values for the simulated clusters, with thick and thin error bars
indicating the 16th-84th percentiles and minimum-maximum values,
respectively. We notice that the low and high-resolution DM-only
simulations provide a comparable amount of subhalos, indicating that
for massive subhalos resolution is not affecting the results, and the
different estimates provided by the CSF and AGN sets are due to the
inclusion of baryonic physics. We verified that the number of subhalos
with circular velocity below the values probed in our analysis is
enhanced in the hydrodynamical runs.

The semi-analytic models predict an amount of massive subhalos similar
to that of N-body and hydrodynamical simulations, suggesting that
orphan galaxies do not contribute considerably to the population of
subhalos at the high-mass end.

For circular velocities in the range $200-400\, {\rm km\,s^{-1}}$, we
find a statistically significant difference between the values of
measured velocity of cluster members in A2142 and those of simulated
subhalos. Simulated clusters present fewer subhalos, with a
discrepancy that is at $\gtrsim 7$ sigma significance level (13.0,
11.9, 6.9, 12.7, 11.6, 10.0 significance level for, respectively,
DMHR, DMLR, CSF, AGN, DLB07 and HEN15 sets. These values become 11.3,
10.0, 7.6, 12.3, 9.5, 8.8 when restricting the analysis to the inner 1
Mpc). We remark that this result is robust, as we are comparing the
outcomes of simulations and direct measurements of velocities, without
any intermediate mass calibration (as done for strong lensing
analyses). The result is even more striking if one considers that it
is just a lower limit. In fact the members with spectroscopic velocity
dispersions are just a fraction of the entire population of galaxy
members. This is shown by the other two histograms that include an
estimate of the circular velocity of elliptical members that have no
SDSS spectroscopic information and then also take into account the
completeness of the sample. The discrepancy is exacerbated by the fact
that we only consider ETGs in A2142, while we consider the whole
population of subhalos in simulated clusters.

\begin{figure}
  \includegraphics[width=\columnwidth]{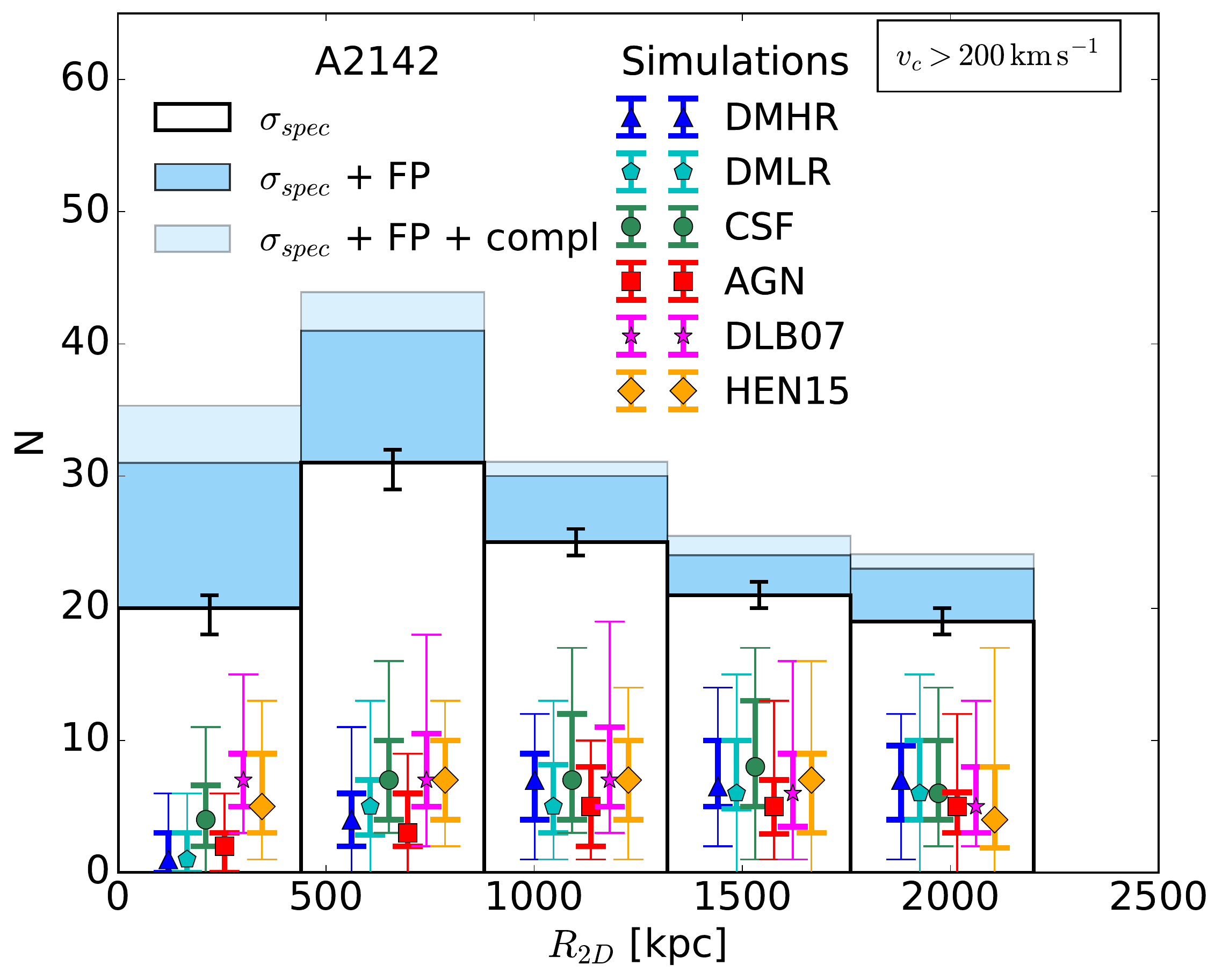}
  \caption{\label{fig: NofR} Projected radial distributions of the
    galaxies shown in Figure \ref{fig: NofV}, with the same
    color-coding. The analysis is restricted to the galaxies having
    circular velocity values larger than $200\, {\rm km\,s^{-1}}$.}
\end{figure}

Finally, we check whether the missing simulated subhalos are
preferentially located at some particular distance from the cluster
center. In Figure \ref{fig: NofR}, we plot the radial distribution of
the galaxies shown in Figure \ref{fig: NofV}, restricting our analysis
to galaxies with circular velocity values larger than $200\, {\rm
  km\,s^{-1}}$. Our results do not depend on the distance from the
cluster center.

To check the robustness of our results, we fit the magnitude-circular
velocity relation and find that the adopted magnitude limit $R_{\rm
  JC} > 20.5$ of O11 corresponds to a circular velocity of
approximately $20\,\rm km\,s^{-1}$, well below the lower limit of
$200\,\rm km\,s^{-1}$ used in our analysis.

\section{Conclusions}
\label{sect: conclusions}
Our analysis indicates that current numerical simulations predict a
significant smaller amount of massive (circular velocity above
$200\,{\rm km\, s^{-1}}$) subhalos. This result is robust, as it holds
even when we compare the predictions of simulations and the direct
measurements of velocity values of cluster members, without addressing
incompleteness issues. When accounting for the latter, the actual
number of observed galaxies becomes larger, making the discrepancy
even more significant. These results support the findings of a recent
strong lensing study of the Hubble Frontier Fields galaxy cluster MACS
J0416 at $z=0.4$ \citep{grillo2015}, suggesting that this discrepancy,
which is already present in DM-only simulations, is not alleviated by
the inclusion of baryonic physics.

\acknowledgments Acknowledgments: E.M. thanks Susana Planelles and
Giuseppe Murante for helpful discussions. E.M. and C.G. acknowledge
support by VILLUM FONDEN Young Investigator Programme through grant
No. 10123. We made use of the CINECA facility PICO, obtained thanks to
a class C ISCRA grant. Financial support for this work was provided in
part by the PRIN-INAF 2014 1.05.01.94.02. The Millennium Simulation
databases and web application were constructed as part of the
activities of the German Astrophysical Virtual Observatory (GAVO).





\begin{thebibliography}{}
\bibitem[Akamatsu et al.(2011)]{akamatsu11} Akamatsu, H., Hoshino, A., Ishisaki, Y., et al.\ 2011, \pasj, 63, S1019
\bibitem[Auger et al.(2010)]{auger2010} Auger, M.~W., Treu, T., Bolton, A.~S., et al.\ 2010, \apj, 724, 511 
\bibitem[Barnab{\`e} et al.(2011)]{barnabe2011} Barnab{\`e}, M., Czoske, O., Koopmans, L.~V.~E., Treu, T., \& Bolton, A.~S.\ 2011, \mnras, 415, 2215 
\bibitem[Bernardi et al.(2003)]{bernardi2003} Bernardi, M., Sheth, R.~K., Annis, J., et al.\ 2003, \aj, 125, 1866 
\bibitem[Biviano(2008)]{biviano08} Biviano, A.\ 2008, arXiv:0811.3535 
\bibitem[Boylan-Kolchin et al.(2008)]{BK2008} Boylan-Kolchin, M., Ma, C.-P., \& Quataert, E.\ 2008, \mnras, 383, 93 
\bibitem[Br{\"u}ggen \& De Lucia(2008)]{bruggen2008} Br{\"u}ggen, M., \& De Lucia, G.\ 2008, \mnras, 383, 1336 
\bibitem[Bruzual \& Charlot(2003)]{bruzual03} Bruzual, G., \& Charlot, S.\ 2003, \mnras, 344, 1000 
\bibitem[Cappellari et al.(2013)]{cappellari2013} Cappellari, M., Scott, N., Alatalo, K., et al.\ 2013, \mnras, 432, 1709 
\bibitem[Cappellari et al.(2015)]{cappellari2015} Cappellari, M., Romanowsky, A.~J., Brodie, J.~P., et al.\ 2015, \apjl, 804, L21 
\bibitem[Chabrier(2003)]{chabrier03} Chabrier, G.\ 2003, \pasp, 115, 763 
\bibitem[Chandrasekhar(1943)]{chandrasekhar1943} Chandrasekhar, S.\ 1943, \apj, 97, 255 
\bibitem[Chen et al.(2012)]{chen12} Chen, Y.-M., Kauffmann, G., Tremonti, C.~A., et al.\ 2012, \mnras, 421, 314 
\bibitem[De Grandi \& Molendi(2002)]{degrandi02} De Grandi, S., \& Molendi, S.\ 2002, \apj, 567, 163 
\bibitem[De Lucia \& Blaizot(2007)]{DLB07} De Lucia, G., \& Blaizot, J.\ 2007, \mnras, 375, 2 
\bibitem[De Lucia et al.(2012)]{delucia2012} De Lucia, G., Weinmann, S., Poggianti, B.~M., Arag{\'o}n-Salamanca, A., \& Zaritsky, D.\ 2012, \mnras, 423, 1277 
\bibitem[Djorgovski \& Davis(1987)]{djorgovski1987} Djorgovski, S., \& Davis, M.\ 1987, \apj, 313, 59 
\bibitem[Dolag et al.(2009)]{dolag2009} Dolag, K., Borgani, S., Murante, G., \& Springel, V.\ 2009, \mnras, 399, 497 
\bibitem[Dressler et al.(1987)]{dressler1987} Dressler, A., Lynden-Bell, D., Burstein, D., et al.\ 1987, \apj, 313, 42 
\bibitem[Dutton \& Treu(2014)]{dutton2014} Dutton, A.~A., \& Treu, T.\ 2014, \mnras, 438, 3594 
\bibitem[Gavazzi et al.(2007)]{gavazzi2007} Gavazzi, R., Treu, T., Rhodes, J.~D., et al.\ 2007, \apj, 667, 176 
\bibitem[Grillo et al.(2008)]{grillo2008} Grillo, C., Lombardi, M., \& Bertin, G.\ 2008, \aap, 477, 397 
\bibitem[Grillo(2010)]{grillo2010} Grillo, C.\ 2010, \apj, 722, 779 
\bibitem[Grillo et al.(2015)]{grillo2015} Grillo, C., Suyu, S.~H., Rosati, P., et al.\ 2015, \apj, 800, 38 
\bibitem[Han et al.(2016)]{han16} Han, J., Cole, S., Frenk, C.~S., \& Jing, Y.\ 2016, \mnras, 457, 1208 
\bibitem[Henriques et al.(2015)]{HEN15} Henriques, B.~M.~B., White, S.~D.~M., Thomas, P.~A., et al.\ 2015, \mnras, 451, 2663 
\bibitem[Humphrey \& Buote(2010)]{humphrey2010} Humphrey, P.~J., \& Buote, D.~A.\ 2010, \mnras, 403, 2143 
\bibitem[Jaffe(1983)]{jaffe1983} Jaffe, W.\ 1983, \mnras, 202, 995 
\bibitem[Jorgensen et al.(1995)]{jorgensen95} Jorgensen, I., Franx, M., \& Kjaergaard, P.\ 1995, \mnras, 276, 1341 \bibitem[Kroupa(2001)]{kroupa01} Kroupa, P.\ 2001, \mnras, 322, 231 
\bibitem[Kravtsov et al.(2004)]{kravtsov2004} Kravtsov, A.~V., Gnedin, O.~Y., \& Klypin, A.~A.\ 2004, \apj, 609, 482 
\bibitem[Markevitch et al.(2000)]{markevitch00} Markevitch, M., Ponman, T.~J., Nulsen, P.~E.~J., et al.\ 2000, \apj, 541, 542 
\bibitem[Moran et al.(2007)]{moran2007} Moran, S.~M., Ellis, R.~S., Treu, T., et al.\ 2007, \apj, 671, 1503 
\bibitem[Munari et al.(2014)]{munari14} Munari, E., Biviano, A., \& Mamon, G.~A.\ 2014, \aap, 566, A68 
\bibitem[Navarro et al.(1996)]{navarro96} Navarro, J.~F., Frenk, C.~S., \& White, S.~D.~M.\ 1996, \apj, 462, 563 
\bibitem[Navarro et al.(1997)]{navarro97} Navarro, J.~F., Frenk, C.~S., \& White, S.~D.~M.\ 1997, \apj, 490, 493 
\bibitem[Okabe \& Umetsu(2008)]{okabe08} Okabe, N., \& Umetsu, K.\ 2008, \pasj, 60, 345 
\bibitem[Owers et al.(2011)]{owers11} Owers, M.~S., Nulsen, P.~E.~J., \& Couch, W.~J.\ 2011, \apj, 741, 122 
\bibitem[Rasia et al.(2015)]{rasia15} Rasia, E., Borgani, S., Murante, G., et al.\ 2015, \apjl, 813, L17 
\bibitem[Remus et al.(2013)]{remus2013} Remus, R.-S., Burkert, A., Dolag, K., et al.\ 2013, \apj, 766, 71 
\bibitem[Rossetti et al.(2013)]{rossetti13} Rossetti, M., Eckert, D., De Grandi, S., et al.\ 2013, \aap, 556, A44 
\bibitem[Saglia et al.(1992)]{saglia1992} Saglia, R.~P., Bertin, G., \& Stiavelli, M.\ 1992, \apj, 384, 433 
\bibitem[Springel et al.(2001)]{springel2001} Springel, V., Yoshida, N., \& White, S.~D.~M.\ 2001, \na, 6, 79 
\bibitem[Springel(2005)]{springel2005} Springel, V.\ 2005, \mnras, 364, 1105 
\bibitem[Steinborn et al.(2015)]{steinborn15} Steinborn, L.~K., Dolag, K., Hirschmann, M., Prieto, M.~A., \& Remus, R.-S.\ 2015, \mnras, 448, 1504 
\bibitem[Thomas et al.(2007)]{thomas2007} Thomas, J., Saglia, R.~P., Bender, R., et al.\ 2007, \mnras, 382, 657
\bibitem[Tornatore et al.(2007)]{tornatore07} Tornatore, L., Borgani, S., Dolag, K., \& Matteucci, F.\ 2007, \mnras, 382, 1050
\bibitem[Treu et al.(2006)]{treu2006} Treu, T., Koopmans, L.~V., Bolton, A.~S., Burles, S., \& Moustakas, L.~A.\ 2006, \apj, 640, 662 
\bibitem[Umetsu et al.(2009)]{umetsu09} Umetsu, K., Birkinshaw, M., Liu, G.-C., et al.\ 2009, \apj, 694, 1643 



\end{thebibliography}
\end{document}